\documentclass[default]{svmult}

\usepackage{mathptmx}       
\usepackage{helvet}         
\usepackage{courier}        
\usepackage{type1cm}        
\usepackage{makeidx}         
\usepackage{graphicx}        
\usepackage{multicol}        
\usepackage[bottom]{footmisc}

\makeindex             

\begin{document}
\title*{Quantum nanomagnets and nuclear spins: an overview}

\author{Andrea Morello}

\institute{Andrea Morello \at Department of Physics and Astronomy,
University of British Columbia, Vancouver, B.C. V6T 1Z1, Canada
\\
Australian Research Council Centre of Excellence for Quantum
Computer Technology, University of New South Wales, Sydney NSW
2052, Australia \\ \email{a.morello@unsw.edu.au}}

\maketitle

\abstract*{This mini-review presents a simple and accessible
summary on the fascinating physics of quantum nanomagnets coupled
to a nuclear spin bath. These chemically synthesized systems are
an ideal test ground for the theories of decoherence in mesoscopic
quantum degrees of freedom, when the coupling to the environment
is local and not small. We shall focus here on the most striking
quantum phenomenon that occurs in such nanomagnets, namely the
tunneling of their giant spin through a high anisotropy barrier.
It will be shown that perturbative treatments must be discarded,
and replaced by a more sophisticated formalism where the dynamics
of the nanomagnet and the nuclei that couple to it are treated
together from the beginning. After a critical review of the
theoretical predictions and their experimental verification, we
continue with a set of experimental results that challenge our
present understanding, and outline the importance of filling also
this last gap in the theory}

\section{Introduction}
\label{intro}

\begin{figure*}[t]
\includegraphics[width=11.7cm]{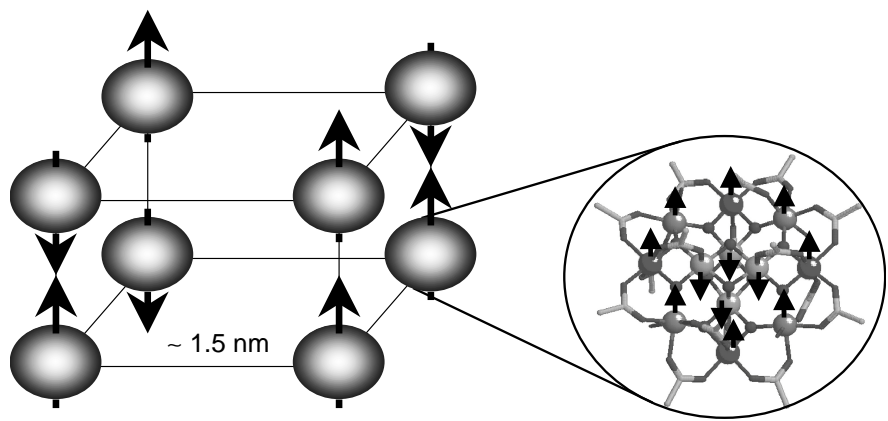}
\caption{\label{structure} Crystalline and molecular structure of a prototypical quantum nanomagnet, Mn$_{12}$-ac
\cite{lis80AC,sessoli93JACS}. The core contains 8 Mn$^{3+}$ (s=2) and 4 Mn$^{4+}$ (S=3/2) ions, which interact by
superexchange to form a ferrimagnetic ground state with total spin S=10. The molecules crystallize in a
tetragonal structure.}
\end{figure*}

In the vast and diverse field of quantum magnetism, the quantum properties of large magnetic molecules have
enjoyed a strong and motivated research activity for more than a decade. Physicists and chemists, theoreticians
and experimentalists, engineers and philosophers, all would find at least one good reason to be interested in
these systems. We shall refer here to ``quantum nanomagnets'' as the broad family of molecules containing a core
of magnetic transition-metal ions, which interact by strong superexchange and which possess magnetic anisotropy
due to crystal field effects \cite{gatteschi94S}. At sufficiently low temperatures (typically $\sim 10$ K), i.e.
much lower than the typical intramolecular exchange interaction energy, the whole molecule effectively behaves as
a nanometer-sized magnet. The total ground state spin value can be rather large, $S \sim 10$, and is often called
``giant spin''. The best studied examples are Mn$_{12}$ \cite{lis80AC}, Fe$_8$ \cite{wieghardt84AC} and Mn$_4$
\cite{aubin96JACS}. The magnetic core of each molecule is stabilized by organic ligands, and the molecules are
typically bound to each other by van der Waals forces to form electrically insulating crystals (Fig.
\ref{structure}). With just a few exceptions \cite{wernsdorfer02N,hill03S}, the magnetic interaction between
different molecules is only of dipolar origin, thus orders of magnitude weaker than the intramolecular exchange.
Single-ion anisotropies, exchange interactions, point symmetry and crystal structure all contribute in a
complicated way to the total magnetic anisotropy of the giant spin. The first great success of molecular magnets
consisted in the demonstration of magnetic bistability and hysteresis \emph{at the molecular level}
\cite{sessoli93N}, i.e. not arising from long-range interactions but only from local ones. Molecules possessing
this sort of bistability were named ``Single Molecule Magnets'' (SMMs). One could then think of 2D arrays of such
molecules \cite{nait-abdi04JAP} as the ultimate magnetic recording medium \cite{joachim00N}, with information
density $\sim 10^{12}$ bit/cm$^2$. For this purpose, the main research goal is to achieve the highest possible
single-molecule anisotropy, to allow magnetic bistability up to - ideally! - room temperature.

However, a proper description of bistability in nanomagnets would be incomplete without quantum mechanics. Since
the anisotropy barrier for reorientation of the giant spin is large but not infinite, there is in principle a
non-vanishing probability for the spin to invert its direction by quantum-mechanical tunneling \emph{through the
barrier} \cite{chudnovsky88PRL}. That is, the magnetic memory would delete itself because of quantum mechanics!
The tunneling probability can be estimated with the knowledge of the anisotropy parameters of the giant spin,
obtained e.g. by electron paramagnetic resonance or neutron scattering. The resulting tunneling rate turns out to
be extremely sensitive to the external magnetic fields applied to the molecule. In Mn$_{12}$ for instance, one
would naively expect the tunneling rate to become astronomically long if a stray field of just $10^{-9}$ T is
applied along the anisotropy axis. The experimental observation of quantum tunneling of magnetization in
Mn$_{12}$ \cite{thomas96N,friedman96PRL,hernandez96EPL,gatteschi03AC} represented a major breakthrough, but in
some sense a puzzling one.

The puzzle was solved by carefully considering the role played by
the coupling between giant spin and surrounding nuclei, which are
always present in the ligands ($^1$H, $^{13}$C, $^{35}$Cl, \ldots)
or in the magnetic ions themselves ($^{55}$Mn, $^{57}$Fe,
$^{53}$Cr, \ldots). The \emph{dynamics} of the nuclear spins
generates a fluctuating magnetic field on the giant spin, thereby
sweeping its energy levels through the tunneling resonance and
yielding a finite tunneling probability \cite{prokof'ev96JLTP}.
Once the giant spin is allowed to have (quantum) fluctuations of
its own, it exerts a back-action on the nuclei and essentially
determines their dynamics. Thus, the system of giant spin + nuclei
\emph{must} be analyzed as a whole, and the resulting theoretical
description is now known as ``theory of the spin bath''
\cite{prokof'ev00RPP}. It predicts, among other things, a
square-root law for the relaxation of magnetization at very
low-$T$ \cite{prokof'ev98PRL}, and a dependence of the tunneling
rate on the isotopic composition of the sample. These predictions
have been promptly verified by measuring the electron spin
relaxation in Fe$_8$ crystals
\cite{wernsdorfer99PRL,wernsdorfer00PRL}, where it is possible to
obtain samples that have stronger (by $^{57}$Fe enrichment) or
weaker (by replacing $^1$H with $^2$H) hyperfine couplings as
compared to those with natural isotopic abundance, while leaving
the giant spin unchanged. The picture has been completed by
experiments looking at the ``other side of the coin'', i.e.
studying directly the dynamics of the nuclear spins by NMR
experiments
\cite{jang00PRL,furukawa01PRB,goto03PRB,morello07CM,morello04PRL,baek05PRB},
and finding that such dynamics is indeed profoundly entangled with
the quantum fluctuations of the giant spin. At the present stage,
much of the efforts in the field are directed towards using
quantum nanomagnets as spin qubits for quantum information
purposes, by pushing the giant spins into a regime where the
tunneling can be made coherent \cite{morello06PRL}.

The purpose of this mini-review is to give an accessible and somewhat pedagogical introduction to the crucial
aspects of the coupled system ``quantum nanomagnet + nuclear spins", what makes it special, what has been
understood, and what requires further attention. It will be shown that much of the common knowledge on nuclear
and electron spin dynamics is entirely inappropriate to describe this system. Importantly, the same often applies
to other quantum degrees of freedom in mesoscopic physics (SQUIDs, quantum dots, \ldots). The discussion given
here on quantum nanomagnets should thus be taken as the ``worked-out example'' of a problem of much broader
interest. The reason for choosing this specific example lies in the wonderful property of quantum nanomagnets to
combine mesoscopic size and fascinating physics, with the cleanliness and reproducibility of a product of
synthetic chemistry.

\section{Theoretical framework}
\label{theory}

\begin{figure*}[t]
\includegraphics[width=11.7cm]{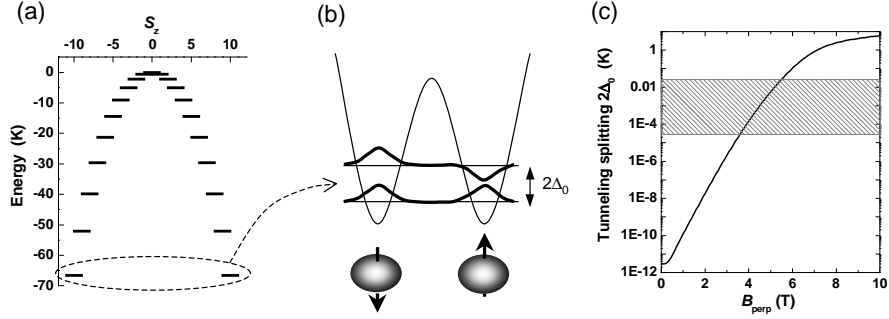}
\caption{\label{levels} (a) Energy levels scheme for the giant spin of Mn$_{12}$-ac in zero external field,
\emph{neglecting} non-diagonal terms in the effective spin Hamiltonian. (b) Including the non-diagonal terms, the
exact eigenstates become symmetric and antisymmetric superpositions of the ``classical'' localized states,
separated by a tunneling splitting $2\Delta_0$. (c) Tunneling splitting as a function of external field applied
perpendicular to the easy axis of anisotropy for Mn${12}$-ac. The hatched area represents the typical range of
hyperfine couplings.}
\end{figure*}

At temperatures such that the internal magnetic excitations can be neglected, i.e. when the entire molecule
behaves as a giant spin of value $S$, the minimal effective spin Hamiltonian that describes the quantum
nanomagnet coupled to a bath of nuclear spins $\{I_k\}$ is:
\begin{eqnarray}
\mathcal{H} = - D S_z^2 + E(S_x^2 - S_y^2) - g \mu_{\rm B}
\mathbf{S}\cdot \mathbf{B} - \sum_k \mathbf{I}_k \tilde{A}_k
\mathbf{S}, \label{Heff}
\end{eqnarray}
where $\hat{z}$ is the easy axis of magnetization. The first term of this Hamiltonian, $-DS_z^2$, gives rise to
an energy levels structure as shown in Fig. \ref{levels}(a), with a parabolic anisotropy barrier separating pairs
of degenerate states. Were this the only term in the Hamiltonian (\ref{Heff}), its eigenstates would be the
eigenstates of $S_z$, i.e. the $|m\rangle$ projections of the spin along the easy axis.

The second term in (\ref{Heff}), $E(S_x^2 - S_y^2)$, is the lowest-order anisotropy term that can break the
uniaxial symmetry of the molecule, and represents a rhombic distortion with hard axis $\hat{x}$. The Hamiltonian
including this term no longer commutes with $S_z$, and its eigenstates are now symmetric and antisymmetric
superpositions of the $|+m\rangle, |-m\rangle$ states. At very low temperatures, such that only the two
lowest-energy states are thermally occupied, the giant spin can be effectively described as a two-level system
(that is, a qubit) with eigenstates $|\mathcal{S}\rangle, |\mathcal{A}\rangle$ separated by a tunneling splitting
$2\Delta_0$ [Fig. \ref{levels}(b)]. A giant spin prepared in one of the ``classical states''
$|\mathcal{Z}^{\pm}\rangle = 2^{-1/2}(|\mathcal{S}\rangle \pm |\mathcal{A}\rangle)$, corresponding to the
magnetization pointing along $\pm \hat{z}$, would tunnel between $|+\mathcal{Z}\rangle$ and
$|-\mathcal{Z}\rangle$ at a frequency $\hbar/2\Delta_0$ \emph{in the absence of external fields}.

The third term, $- g \mu_{\rm B} \mathbf{S}\cdot \mathbf{B}$,
describes the coupling to a magnetic field. If $\mathbf{B} \perp
\hat{z}$, this represents an extra non-diagonal term which has the
effect of rapidly increasing the tunneling splitting
$2\Delta_0(B_{\perp})$ [Fig. \ref{levels}(c)]. Conversely,
$\mathbf{B}
\parallel \hat{z}$ has the effect of breaking the degeneracy of the ``classical'' states $\pm |m\rangle$. If the longitudinal bias $\xi = g
\mu_{\rm B} S_z B_z$ is much larger than $2\Delta_0$, the spin is
effectively localized on one side of the barrier, with vanishing
probability to tunnel.

Finally, the term $- \sum_k \mathbf{I}_k \tilde{A}_k \mathbf{S}$
represents the coupling to the nuclear spin bath. Although in some
instances the coupling tensor $\tilde{A}$ may be isotropic (e.g.
for the $^{55}$Mn nuclei in Mn$^{4+}$ ions), this is not true in
general. Also, strictly speaking an external field acts on the
nuclei as well with a term $-\gamma_k \mathbf{B}\cdot
\mathbf{I}_k$ ($\gamma$ is the nuclear gyromagnetic ratio), but
this can be usually neglected in comparison with the hyperfine
coupling. If one were to assume that the nuclear spins are
\emph{static}, then the effect of the hyperfine field on the giant
spin would be the same as that of a static external field, with
typical strength $10^{-5} - 10^{-3}$ T, or $0.1 - 10$ mK in terms
of coupling energy. The spread of possible hyperfine couplings is
incorporated in a parameter $E_0$. By inspecting Fig.
\ref{levels}(c) for the case $\mathbf{B}=0$, we see that a
transverse component of the hyperfine field would have hardly any
effect on the nanomagnet's tunneling splitting, whereas a
longitudinal component gives a bias $\xi_{\rm N}$ that is easily
several orders of magnitude larger than $2\Delta_0$. Under these
conditions, the giant spin should be frozen for the eternity. The
case could be made even stronger by accounting also for the
dipolar interaction between molecules in the crystal, which adds
an extra (static) random field of order $10^{-3} - 10^{-2}$ T. So
why do we observe tunneling in the experiments?

The answer comes from the \emph{dynamics} of the nuclear spins. At the very least, the nuclear spin bath will
have an intrinsic dynamics due to the mutual dipolar couplings, which can be described by an additional term of
the form $\sum_{k,k'} \mathbf{I}_k \tilde{V}_{k,k'} \mathbf{I}_{k'}$. This effectively generates a bias on the
giant spin that fluctuates on a timescale of the order of the nuclear $T_2$ ($\sim 1 - 10$ ms typically). The
amplitude of this fluctuation can be sufficient to sweep the electron spin levels through the tunneling
resonance, for that tiny minority of molecules that finds itself having $\xi \simeq 0$ at some time. The sweeping
of hyperfine bias through the resonance yields a Landau-Zener - like incoherent tunneling probability. Once a
giant spin has tunnelled, two crucial things happen: (i) the distribution of internal dipolar fields in the
crystal suddenly changes, giving other molecules a chance to have $\xi \simeq 0$, etc\ldots; (ii) the hyperfine
field on the nuclear spins belonging to or surrounding the tunnelled molecule suddenly changes, stimulating
further dynamics of the nuclear spins. The consequences of (i) are the formation of a ``hole'' in the
distribution of dipolar biases \cite{tupitsyn04PRB}, with width related to the spread of hyperfine bias, and a
$\propto \sqrt{t}$ law for the short-term relaxation of the magnetization \cite{prokof'ev98PRL}. Here we
concentrate on the aspect (ii), namely what happens to the nuclear spins as a consequence of a tunneling event.
This point is extremely interesting and instructive, because it radically deviates from the framework under which
the dynamics of nuclear spins is commonly analyzed \cite{abragam61,slichterB}.

Practically all of the theory of nuclear magnetism is based on perturbation theory, since one typically has a
large \emph{static} magnetic field, $B_z$, applied from the outside, plus some local fluctuations of much smaller
amplitude. For instance, one can estimate the longitudinal nuclear relaxation rate, $T_1^{-1}$, by evaluating the
magnitude of the transverse component of the fluctuating field, $b_{\perp}$, and assuming that the
time-correlation of the fluctuations decays exponentially with time constant $\tau$, $\langle b_{\perp}(t_1)
b_{\perp}(t_2)\rangle \simeq \langle b_{\perp}^2 \rangle \exp(-(t_2 - t_1)/\tau)$. Within perturbation theory,
this yields:
\begin{eqnarray}
\frac{1}{T_1} \simeq \gamma^2 \langle b_{\perp}^2 \rangle
\frac{\tau}{1 + \omega_{\rm N}^2 \tau^2}, \label{nslr}
\end{eqnarray}
with $\gamma$ and $\omega_{\rm N}$ the nuclear gyromagnetic ratio
and Larmor frequency, respectively. Eq. (\ref{nslr}) simply means
that the nuclear relaxation rate is proportional to the power
spectrum of the fluctuating field, taken at the nuclear Larmor
frequency. The reader shall recognize that this is just the
fluctuation-dissipation theorem, a well-known result of linear
response theory \cite{kuboB}. Expressions like (\ref{nslr}) are
ubiquitous in the NMR literature, since they relate an
experimentally accessible quantity (the nuclear $T_1$) to the
dynamic properties of the environment where the nuclei are
immersed ($\tau$), thereby giving NMR its status of ``local
probe'' for the dynamics of complex systems. However, \emph{the
nuclear spin dynamics in tunneling nanomagnets is one example
where the use of expressions derived from perturbation and linear
response theory is incorrect and unjustified}. This is
particularly true for the nuclei that ``belong'' to a magnetic ion
in the molecule core, like $^{55}$Mn or $^{57}$Fe.

Perturbation theory breaks down here because the hyperfine field produced by the electron spins on their nuclei
(typically $\sim 30$ T in Mn and $\sim 50$ T in Fe) is much larger than any externally applicable field.
Therefore, when a spin tunneling event occurs, the nuclear Hamiltonian suddenly changes by a large amount, and
the effect of the ``fluctuation'' of the electron spin cannot be treated as a perturbation. In addition, since
the electron spin tunneling events are allowed in the first place by nuclear spin fluctuations (at least in the
small transverse field limit), it is not even possible to treat the nanomagnet as an independent source of
``field jumps''. The dynamics of the nanomagnet and its nuclei must be treated together, since the one drives the
other and vice versa: again, this is a very uncommon situation in NMR.

The problem is treated formally in the spin bath theory by writing a master equation for the probability
$P_{\Uparrow \Uparrow}(t)$ for the giant spin to remain in the $|\Uparrow\rangle$ until time $t$
\cite{prokof'ev95CM}. For a nanomagnet alone, that would be trivially:

\begin{eqnarray}
P_{\Uparrow \Uparrow}(t) = 1 - \frac{\Delta_0^2}{\epsilon^2} \sin^2{\epsilon t}, \label{Pupup} \\
\epsilon = \sqrt{\xi^2 + \Delta_0^2}. \label{epsilon}
\end{eqnarray}

When coupled to the nuclear spin bath, the tunneling of
$\mathbf{S}$ has the effect of ``coflipping'' some of the nuclei.
Given the details of the spin-bath coupling, one can calculate
what is the ``natural'' number of nuclei that would flip upon a
tunneling event, by considering two effects. (i) If the hyperfine
field acting on a nuclear spin does not exactly invert its
direction by $180^{\rm o}$ upon tunneling (which can happen if the
nucleus is subject to dipolar fields from different nanomagnets,
only one of which tunnels at some instant, or if an external
transverse field is applied), then the quantization axis of the
nucleus changes, i.e. the spin precesses around a new axis.
Quantum mechanically, this is equivalent to a flip of the nuclear
spin. This mechanism depends only on the direction and not on the
timescale of the hyperfine field jumps. (ii) Even if the hyperfine
field changes direction by exactly $180^{\rm o}$, the nuclear spin
may still follow if $\mathbf{S}$ flips slowly. The nuclear
coflipping probability is then proportional to $(\omega_{\rm
N}/\Omega_0)^2$, where $\Omega_0$ is the ``bounce frequency'' of
the nanomagnet, i.e. the frequency of the oscillations of
$\mathbf{S}$ on the bottom of each potential well. However, since
typically $\Omega_0 \sim 10^{12}$ s$^{-1}$, this mechanism is
usually unimportant. Knowing the strength and direction of the
individual hyperfine couplings, one can calculate the average
number, $\lambda$, of nuclear spins that would coflip with
$\mathbf{S}$ at each tunneling event. Given a certain arbitrary
state of the nuclear spin bath at some instant, a tunneling event
may require a number $M$ of nuclei to coflip with $\mathbf{S}$ in
order to conserve energy. If $M \gg \lambda$, such an event is
essentially forbidden (``orthogonality blocking''
\cite{prokof'ev95CM}). This is accounted for in the theory by
writing an effective tunneling matrix element, $\Delta_M \simeq
\Delta_0 \lambda^{M/2} / M!$, which has its maximum value for
$M=0$ (and coincides with the half-tunneling splitting of the
nanomagnet alone), and goes quickly to zero for $M \gg \lambda$.
Thus $P_{\Uparrow \Uparrow}(t)$ may take any of the possible
values $P_M (t)$ obtained by replacing $\Delta_M$ for $\Delta_0$
in Eq. (3). In addition, since $P_M (t)$ depends on the bias $\xi$
[Eq. (\ref{epsilon})], which has a hyperfine component $\xi_{\rm
N}$ that can fluctuate over a range $E_0$, one must also average
$P_M (t;\xi)$ over the bias distribution, obtaining:

\begin{eqnarray}
\langle P_M(t;\xi)\rangle_{\xi} - \frac{1}{2} = \frac{1}{2} \exp(-\Gamma_M^{\rm N} t),\\
\Gamma_M^{\rm N} = \frac{2\Delta_M^2}{\sqrt{\pi}\hbar E_0},
\end{eqnarray}

where $\Gamma_M^{\rm N}$ represents the rate for $\mathbf{S}$ to
tunnel accompanied by the coflip of $M$ nuclear spins. Calling
$\xi_0$ the energy scale associated with the flip of $\lambda$
nuclei (the ``natural'' number of coflips), and considering that
the highest tunneling probability is obtained for $M=0$, the
leading term for the global tunneling rate becomes:
\begin{eqnarray}
\Gamma^{\rm N} \simeq \frac{2\Delta_0^2}{\sqrt{\pi}\hbar E_0}
\exp(-|\xi_{\rm B}|/\xi_0), \label{rate}
\end{eqnarray}
where $\xi_{\rm B} = g \mu_{\rm B} S_z B_{\rm tot}$ is the static
component of the bias. $B_{\rm tot}$ here is the sum of the
longitudinal components of an externally applied field and of the
dipolar field from neighboring nanomagnets. Since the latter is
itself time-dependent as soon as tunneling events start occurring,
this leads to an interesting collective dynamics, characterized by
a square-root time relaxation \cite{prokof'ev98PRL}. Starting from
a fully magnetized sample, $\mathcal{M}(0)=\mathcal{M}_{\rm sat}$,
the short-time behavior of the magnetization is:

\begin{eqnarray}
\mathcal{M}(t)/\mathcal{M}_{\rm sat} \simeq 1 - \sqrt{t/\tau_{\rm short}}, \label{sqrtrel}\\
\tau_{\rm short}^{-1} \sim \frac{\xi_0}{E_{\rm D}} \Gamma^{\rm
N}(\xi=0), \label{taushort}
\end{eqnarray}

where $E_{\rm D}$ is the spread of dipolar biases. We see
therefore that the microscopic properties of the spin bath enter
directly in the macroscopic relaxation of an ensemble of
nanomagnets, through the strength of the hyperfine couplings
[$E_0$, in Eq. (\ref{rate})] and the coflipping probability
[$\xi_0$, in Eqs. (\ref{rate}),(\ref{taushort})].

From the opposite perspective, we have seen that to each tunneling
event one can associate a certain nuclear coflipping probability.
However, for the purpose of comparing theory and experiments, it
is convenient to translate this into more common NMR language by
calculating the nuclear longitudinal $(T_1)$ and transverse
$(T_2)$ relaxation times. The latter is relatively easy to
estimate because, unless the tunneling rate is made extremely high
by applying a strong transverse field, one will generally have
$T_2^{-1} \ll \Gamma^{\rm N}$, i.e. $\mathbf{S}$ remains static on
the timescale of the transverse nuclear relaxation. On such a
short timescale we therefore recover the applicability of the
standard perturbative treatments, whereby $T_2$ is determined by
the nuclear dipole-dipole couplings and can be calculated using
the van Vleck method \cite{vanvleck48PR,abragam61}. Conversely,
$T_1$ is determined precisely by the tunneling rate of the
nanomagnet, which determines how often the local field on the
nuclear spins changes direction. Quite amusingly, this problem was
first treated by Abragam (Ref. \cite{abragam61}, page 478) as
``\ldots an example that has no physical reality but where the
result can be obtained very simply \ldots''; 40 years later, that
example has found physical reality in quantum nanomagnets! The
simple result is that, since the local field changes direction at
intervals $\sim \tau$, then \cite{abragam61,alexander65PR}
\begin{eqnarray}
T_1^{-1} \sim 2\Gamma^{\rm N}. \label{T1}
\end{eqnarray}
Once again, the point to bear in mind here is that $\Gamma^{\rm
N}$ is itself strongly dependent on the nuclear spin dynamics.
Moreover, one should be careful before calling this the nuclear
spin-\emph{lattice} relaxation rate, as $T_1^{-1}$ is normally
interpreted. We shall come back to this issue in the review of the
experimental results, but here we simply note that no phonon bath
has been introduced so far, in the context of nuclear-spin
mediated tunneling. A phonon-mediated tunneling rate,
$\Gamma^{\phi}$, can also be calculated
\cite{kagan80JETP,stamp04CP}, yielding:
\begin{eqnarray}
\Gamma^{\phi} \simeq \frac{\Delta_0^2 W_{\phi}(T)}{\Delta_0^2 + \xi^2 + \hbar^2 W_{\phi}^2(T)}, \label{Gammaphi}
\end{eqnarray}
where $W_{\phi}$ is the linewidth of the electron spin states due to intra-well phonon-assisted transitions
\cite{leuenberger00PRB}. At low temperatures this tunneling rate is orders of magnitude smaller than the
nuclear-driven one.

\section{Experimental results}
\label{experiment}

\begin{figure*}[t]
\includegraphics[width=11.7cm]{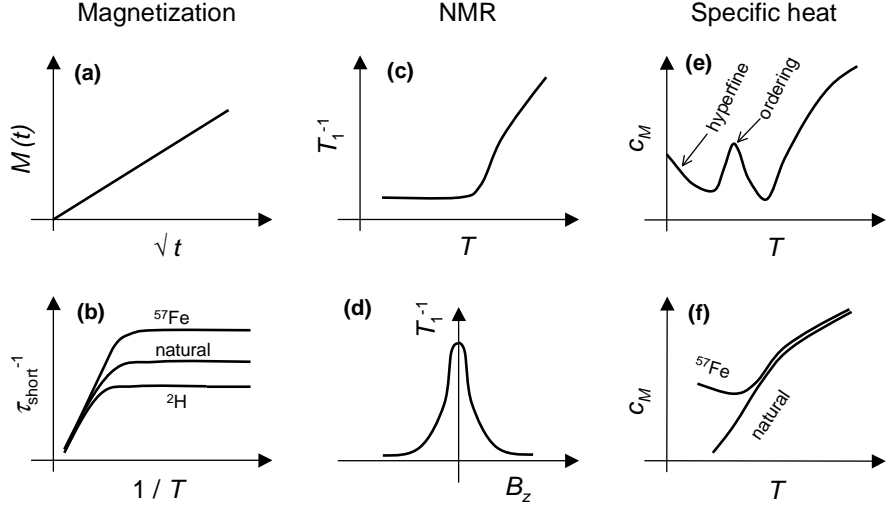}
\caption{\label{expt} Sketches of the qualitative features of some
spin-bath related phenomena in quantum nanomagnets. (a)
Square-root time relaxation of magnetization in Fe$_8$
\cite{wernsdorfer99PRL}, Eq. (\ref{sqrtrel}). (b) Isotope effect
in the short-time magnetic relaxation rate in Fe$_8$, Eq.
(\ref{taushort}), and $T$-independent relaxation at low-$T$
\cite{wernsdorfer00PRL}. (c) Crossover to $T$-independent nuclear
relaxation rate $T_1^{-1}$ driven by quantum tunneling in Fe$_8$
\cite{ueda02PRB,baek05PRB} and Mn$_{12}$-ac \cite{morello04PRL},
Eq. (\ref{rate}). (d) Longitudinal field suppression of quantum
tunneling in Mn$_{12}$-ac as seen by nuclear $T_1^{-1}$
\cite{morello07CM,morello04PRL}. (e) Long-range magnetic ordering
in Mn$_4$ and equilibrium hyperfine specific heat at low-$T$
\cite{evangelisti04PRL}. (f) Isotope effect in the magnetic
specific heat of Fe$_8$ \cite{evangelisti05PRL}.}
\end{figure*}

From the theoretical treatment of the coupled nanomagnet-nuclear
spins system, we have obtained several testable predictions. First
of all, it is predicted that an initially magnetized sample would
relax in a non-exponential way, given by Eq. (\ref{sqrtrel}). This
has been verified very early on by magnetization measurement on
Fe$_8$ at very low temperatures, which show $T$-independent
relaxation below $T\sim 360$ mK, and indeed, a square-root time
relaxation function \cite{wernsdorfer99PRL} [Fig. \ref{expt}(a)].
Second, but perhaps more important for the present discussion, the
nuclear-driven tunneling rate $\Gamma^{\rm N}$ should depend on
the details of the hyperfine couplings, through the spread of
hyperfine bias, $E_0$, and the energy scale associated with the
``natural'' number of nuclear spins that coflip with the
nanomagnet upon each tunneling event, $\xi_0$. Both can be changed
by isotopic substitution of some constituents. Iron-based
nanomagnets, like Fe$_8$, are particularly suited for this type of
study because the strength of the spin-bath couplings can be
either increased by replacing $^{56}$Fe $(I = 0)$ with $^{57}$Fe
($I = 1/2$), or decreased by replacing $^1$H ($I=1/2$) with $^2$H
($I=1$ but 6.5 times smaller gyromagnetic ratio). When repeating
the low-$T$ magnetization relaxation experiments on isotopically
substituted samples, it was found that the relaxation rate would
increase with $^{57}$Fe enrichment, and decrease with $^2$H
substitution \cite{wernsdorfer00PRL} [Fig. \ref{expt}(b)]. Notice
that in both cases the mass of the molecule is increased, so that
no phonon isotope effect could explain the observed trend.
Finally, by measuring the magnetization decay while applying a
special sequence of longitudinal fields, it was confirmed that the
fluctuating hyperfine bias creates a ``tunneling window'' within
which the nanomagnets can undergo quantum relaxation
\cite{wernsdorfer99PRL,wernsdorfer00PRL,tupitsyn04PRB}. The width
of this tunneling window was found to be dependent on the isotopic
composition of the sample. Thus, the effect of nuclear spins on
the quantum tunneling of an ensemble of nanomagnets has been
verified very early on and in a rather uncontroversial way.

The opposite effect, i.e. the influence of quantum tunneling of
the nanomagnets on the nuclear spin dynamics, was observed short
thereafter for $^1$H \cite{ueda02PRB} and $^{57}$Fe in Fe$_8$
\cite{baek05PRB}, and $^{55}$Mn in Mn$_{12}$
\cite{morello07CM,morello04PRL} [Fig. \ref{expt}(c)]. In each
experiment, a $T$-independent nuclear $T_1$ was found below a
certain temperature, and both for Fe$_8$ and Mn$_{12}$ this was
comparable to the temperature at which, for instance, the magnetic
hysteresis loops also became $T$-independent
\cite{wernsdorfer99S,chiorescu00PRL,bokacheva00PRL}. This can be
qualitatively understood from Eq. (\ref{T1}), which obviously
implies that when the $T$-independent $\Gamma^{\rm N}$ is the
dominant electron spin fluctuation rate, also $T_1^{-1}$ should be
$T$-independent. At higher temperatures, thermally-assisted
transitions in the nanomagnet start to play a role, and the
situation becomes much more complicated. This has led to some
controversy in the interpretation of the nuclear relaxation data
in the thermally activated regime, on which we shall not dwell.
Focusing instead on the low-$T$ quantum regime, it's worth noting
that one could look for the other obvious signature of resonant
tunneling, i.e. that it should be strongly suppressed by a
longitudinal external field. Indeed, it has been observed that the
nuclear $T_1$ in Mn$_{12}$ at very low temperatures becomes almost
two orders of magnitude slower with the application of a small
longitudinal field \cite{morello07CM,morello04PRL}, as compared to
the zero-field case [Fig. \ref{expt}(d)]. The interpretation of
the experiments on Mn$_{12}$-ac is slightly complicated by the
fact that this particular compound contains a minority of
fast-relaxing molecules \cite{sun99CC,wernsdorfer99EPL} which
remain dynamic down to the lowest temperatures, whereas the
majority species have a negligibly small tunneling rate in zero
field. On the other hand, the fact that a large fraction of the
sample can remain fully magnetized for long times (months), has
allowed to measure the nuclear dynamics in Mn$_{12}$-ac as a
function of the sample magnetization, and revealed the effect of
dipolar coupling between nuclear spins (i.e. the nuclear spin
diffusion described by the term $\sum_{k,k'} \mathbf{I}_k
\tilde{V}_{k,k'} I_{k'}$ \cite{prokof'ev95CM,prokof'ev00RPP}).
Since spin diffusion can occur only between nuclei subject to the
same local magnetic field, a demagnetized sample where half of the
spin are ``up'' and half of the spins are ``down'' should have
slower spin diffusion by a factor $\sqrt{2}$, as compared to a
fully magnetized sample. This has been indeed verified by
measuring the transverse relaxation rate $T_2^{-1}$
\cite{morello07CM,morello04PRL}.

So far, we have discussed experiments that essentially confirmed
the predictions of the theory of the spin bath. Not surprisingly,
the situation becomes more puzzling when looking at the thermal
properties of quantum nanomagnets and spin bath, because the
theory does not deal with them. Specific heat is obviously the
main experimental tool to investigate thermal equilibrium (or lack
thereof) in an ensemble of nanomagnets. It was found early on that
the magnetic specific heat in Fe$_8$ would be unmeasurable at very
low temperatures and zero external field, because the spin-lattice
relaxation time of the electron spins becomes much longer that the
typical timescale of the experiment ($\sim 10^3$ s at most).
However, by applying a large transverse magnetic field, the
tunneling rate could be made large enough to recover the
equilibrium magnetic specific heat \cite{luis00PRL,mettes01PRB}.
Importantly, the case of Mn$_{12}$-ac is different in that one
could observe the contribution of the nuclear spins, which have a
large specific heat at millikelvin temperatures. The magnetic
specific heat of Mn$_{12}$-ac reveals a hyperfine contribution
which approaches the full equilibrium value when a transverse
field is applied, but remains at least partially visible even in
zero external field. This observation seems to imply that the
nuclear spins in Mn$_{12}$-ac find a way to thermalize to the
phonon bath even at very low-$T$, and even without large
transverse fields. The definitive experimental proof of the
thermal equilibrium in the nuclear spins of Mn$_{12}$-ac was then
found by directly measuring the $^{55}$Mn nuclear spin temperature
by NMR techniques \cite{morello07CM,morello04PRL}. This is a
crucial observation because the nuclear spins have essentially no
direct link to the phonon bath, therefore their thermal
equilibration must proceed via the coupling to the electron spins
and their subsequent spin-lattice relaxation process. However, we
are dealing here with a temperature regime where the only electron
spin transitions are quantum tunneling ones. The thermalization
rate $\tau_{\rm th}^{-1}$ is found to be orders of magnitude
faster than expected from the known phonon-assisted tunneling rate
[Eq. (\ref{Gammaphi})], and is actually very close to the observed
nuclear-spin mediated tunneling rate. The precise value appears to
be a matter of sample size, cooling power and thermal contact to
the refrigerator, and it's quite plausible that $\tau_{\rm
th}^{-1} \simeq \Gamma^{\rm N}$ in the limit of small sample and
perfect contact to the thermal bath \cite{morello07CM}.

If the nuclear spins are found to be in thermal equilibrium, one
may argue that the electron spins should be in thermal equilibrium
as well. Since the latter are mutually coupled by dipolar
interactions of the order $\xi_{\rm D} \sim 0.1$ K, at very
low-$T$ we may expect the ensemble of nanomagnets to undergo a
transition to a magnetically ordered state, provided the timescale
involved in finding the collective ground state is short enough.
Long-range magnetic ordering in molecular magnets was first found
in the Mn$_6$ compound \cite{morello03PRL,morello06PRB}, which has
negligible anisotropy and therefore maintains a very fast electron
spin-lattice relaxation time down to the lowest temperatures. The
electron-spin transitions involved in finding the ordered state
are somewhat trivial (no tunneling), but an additional important
result was the confirmation that the nuclear $T_1$ measured by NMR
precisely coincides with the nuclear spin-\emph{lattice}
relaxation time as measured by specific heat \cite{morello06PRB}.
The first quantum nanomagnet to show long-range magnetic ordering
was a special type of Mn$_{4}$ \cite{evangelisti04PRL} [Fig.
\ref{expt}(e)], characterized by fast tunneling rate in zero
field. In this case, the collective ordered state is found through
quantum relaxation, and the ferromagnetic phase transition causes
a peak in the specific heat at the temperature where the
nanomagnets undergo long-range ordering. Also in this system the
nuclear spins were found to be in thermal equilibrium. That the
nuclear spins should play a role in this process is clear from the
discussions above, and it has been experimentally verified in
Fe$_8$, again by playing with isotopic substitutions. While a
``natural'' Fe$_8$ sample falls out of thermal equilibrium at
low-$T$ in zero field, a sample enriched with $^{57}$Fe shows a
much larger magnetic specific heat that almost reaches the full
equilibrium value \cite{evangelisti05PRL} [Fig. \ref{expt}(f)]. At
the lowest temperatures, the hyperfine contribution to the
specific heat is also revealed. These results prove once again
that the thermalization time, the electron spin tunneling rate and
the nuclear $T_1$ are closely related and essentially belong to
the same physical phenomenon, i.e. the inelastic tunneling of the
nanomagnet spin, driven by the dynamics of the nuclear spin bath,
where the phonon bath acts as a thermostat for both electrons and
nuclei.

\section{Open questions and future directions}
\label{conclusion}

After discussing the experiments involving thermalization of the
nuclei and electron spins, it should be clear that the missing
ingredient in the present description of nanomagnet + spin bath is
the role of the phonon bath in influencing the tunneling
transition \emph{in the presence of a fluctuating hyperfine bias}.
To make the point absolutely clear, let us ask the question: ``How
does a nanomagnet know what is its most energetically favorable
spin direction \emph{at the instant when a tunneling event can
occur}, so that it can participate to a long-range ordered
state?'' In general, each nanomagnet is subject to some bias
originating from the dipolar field of its neighbors, plus the
hyperfine bias from the nuclei. Only when the two compensate each
other, can a tunneling event occur. At that moment, however, the
total energy of the nanomagnet + nuclei is the same regardless of
the direction of the nanomagnet spin. The difference between the
two possible orientations is that one will have the electron spin
in a favorable direction with respect to the local dipolar field,
but the nuclear spins pointing against the hyperfine field, and
vice versa in for the other orientation. It seems that the
creation of long-range order in the nanomagnets should go at the
expenses of the nuclear energy, but we know this is not the case
since both electron and nuclear spins are found to be in thermal
equilibrium down to the lowest temperatures, and the equilibrium
nuclear specific heat is well visible precisely when long-range
magnetic ordering is observed. Thus, the theory needs to be
improved to include the role of phonons in this process
\cite{morello07CM}.

One reason to stress this point is that the attention of the community is progressively being shifted towards
\emph{coherent} tunneling processes in quantum nanomagnets, for the purpose of quantum information processing.
Then we shall be interested in modelling, and ultimately measuring, the decoherence rate of the electron spins
under the most favorable conditions. In particular, it has been shown that when considering the effect of nuclear
spins and phonons on the decoherence rate, one expects a ``coherence window'' where the coupling to the
environment is the least destructive \cite{stamp04PRB}. Crudely speaking, at low tunneling frequencies the
nanomagnet will couple strongly to the nuclear spins, because their Larmor frequencies are comparable, whereas
higher tunneling frequencies would increase the coupling to phonon modes, leaving an optimal low-coupling point
between these regimes. For dense (i.e., undiluted) and crystalline ensembles of nanomagnets, the dipolar
couplings are actually more important than the hyperfine ones, and the (strongly $T$-dependent) optimal operation
point is determined by the crossing between dipolar and phonon decoherence rates \cite{morello06PRL}. However, in
view of the above discussion on the shortcomings of the present description of tunneling in the presence of spin
and phonon baths, one may wonder to what extent are these predictions accurate. But this is precisely the beauty
and the strength of the research on quantum nanomagnets: theoretical predictions can be tested qualitatively and
quantitatively against the experiment, as has been done in the past ten years to understand the incoherent
tunneling regime. It should be possible to single out each contribution to decoherence by virtue of its
dependence on temperature, field, and isotopic substitution. While the experiments needed to demonstrate coherent
control of quantum nanomagnets are extremely demanding, the precious information harvested by studying relaxation
and dephasing in these systems has the potential to push our understanding of nanometer-sized quantum systems to
unprecedented levels.

\begin{acknowledgement}
The author is deeply indebted with many colleagues and coworkers who
have supported, stimulated or challenged his activity in this field
of research. In approximate chronological order: F. Luis, L. J. de
Jongh, R. Sessoli, H. B. Brom, G. Arom\'{i}, P. C. E. Stamp, I. S.
Tupitsyn, B. V. Fine, W. Wernsdorfer, M. Evangelisti, W. N. Hardy,
Z. Salman, R. F. Kiefl, J. C. Baglo, A. L. Burin.
\end{acknowledgement}

\end{document}